\documentclass[cits]{PoS}

\newcommand{\dov}{\ensuremath{D_\textrm{\scriptsize OV}}}
\newcommand{\dci}{\ensuremath{D_\textrm{\scriptsize CI}}}
\newcommand{\dwi}{\ensuremath{D_\textrm{\scriptsize WI}}}

\title{The eigenvalue spectrum for dynamical Chirally Improved fermions}

\ShortTitle{The eigenvalue spectrum for dynamical Chirally Improved fermions}

\author{\speaker{Martina Joergler} and C. B. Lang\\
        Institut f\"ur Physik, FB Theoretische Physik, Universit\"at Graz, 
        A-8010 Graz, Austria\\
        E-mail: \email{martina.joergler@edu.uni-graz.at}, 
        \email{christian.lang@uni-graz.at}}

\abstract{We study the eigenvalues of Dirac operators in QCD with two mass
degenerate dynamical fermions. The gauge configurations have been obtained
with HMC and the so-called Chirally Improved fermionic action. We compare
eigenvalues obtained for the overlap Dirac operator on these configurations
with those for the Chirally Improved (CI) operator (studied earlier). Results
of Random Matrix Theory allow us to determine the chiral condensate.}

\FullConference{The XXV International Symposium on Lattice Field Theory\\
		 July 30 - August 4 2007\\
		 Regensburg, Germany}

\begin{document}

\section{Dirac operators: \dwi,  \dov, and \dci}
In this work we compare the properties of the eigenvalue spectra of three
lattice Dirac operators, namely the Wilson operator \dwi, the overlap operator
\dov\ and the Chirally Improved (CI) operator, denoted as \dci. The well
known \dwi\ (for vanishing mass parameter $m$) is given by
\begin{equation} 
\dwi(m,n) = 
\frac{4}{a}\, \mathbf{1} -  \frac{1}{2\, a} \sum_{\mu = \pm1}^{\pm
4}(\mathbf{1}-\gamma_\mu) \,U_\mu(n) \,\delta_{n+\hat{\mu},m}\;, 
\end{equation}
using the notation $\gamma_{-\mu} = -\gamma_\mu$. The operator violates
chiral symmetry, and thus does not satisfy the Ginsparg-Wilson equation. The
second fermionic action studied here is defined through \dov\
\cite{Neuberger:1997fp}:
\begin{equation} 
\label{dov} 
\dov = \frac{1}{a} \left( \mathbf{1} - 
\gamma_5\;\rm{sign}(\gamma_5 \,A)\right)\;,
\end{equation}
with $A=\mathbf{1}\;s - a\,\dwi$ and $0<s<2$. For our calculations, we chose
$s=1.8$.

This operator is an exact solution of the Ginsparg-Wilson equation
\begin{equation} 
\label{gweq}\gamma_5\, D + D \,\gamma_5 = D \,\gamma_5 \,D\;, 
\end{equation}
and therefore implements chiral symmetry on the lattice. As a consequence of
this the spectrum of \dov\ lies exactly on a circle, the so-called
Ginsparg-Wilson circle.

The third Dirac operator we used, \dci, represents an
approximate solution to (\ref{gweq}). It is defined \cite{Gattringer:2000js}
as a truncated expansion of a most general solution of the Ginsparg-Wilson
equation into 'paths' on the lattice of varying length. Taking paths up to
infinite length results in an exact solution. Using this technique we can
combine lower computer cost with -- approximate, but good -- chiral
properties \cite{BGR04}.

\section{Dynamical Chirally Improved fermions}

For the following analysis of the spectral properties of the Dirac operators
we use gauge fields with 2 dynamical flavors of fermions with degenerate
masses. These were constructed with an HMC-algorithm, implemented with the
L\"uscher-Weisz gauge action and the already mentioned CI fermionic action.
More details can be found in \cite{Lang:2005jz}.

Due to the 'almost chiral' properties of this action, our HMC produces
gauge-configurations which frequently tunnel between  different topological
sectors within one Markov chain. Table \ref{hmc} gives a short summary of the
parameters of the gauge fields used in our analysis.

\begin{table}[ht]
\centering{\small
\begin{tabular}{|c|c|c|c|c|c|c|}
\hline
run & $L^3\times T$ & $\beta_1$ & $am$ & $a[\rm{fm}]$ & $am_{AWI}$& \#confs. \\
\hline
a & $12^3 \times 24$ & $5.2$ &$0.02$ &$0.115(6)$ &$0.025$& $73$ \\
b & $12^3 \times 24$ & $5.2$ &$0.03$ &$0.125(6)$& $0.037$ &$52$ \\
c & $12^3 \times 24$ & $5.3$ &$0.04$ &$0.120(4)$& $0.037$& $55$ \\
d & $12^3 \times 24$ & $5.3$ &$0.05$ &$0.129(1)$ &$0.050$ &$40$ \\
\hline
\end{tabular}}
\caption{The parameters for the runs used for this work. $L^3\times
T$ denotes the extent of the lattice in units of the lattice spacing $a$, $am$
the bare mass parameter of \dci, $am_{AWI}$ the quark mass calculated via
the axial Ward identities, and $\beta_1$ the first gauge coupling of the set
of three LW-couplings used \cite{Lang:2005jz}. The configurations are separated
by 10 HMS-trajectories.}
\label{hmc}
\end{table}

\section{Comparison of the spectral properties} 

\begin{figure}[ht]
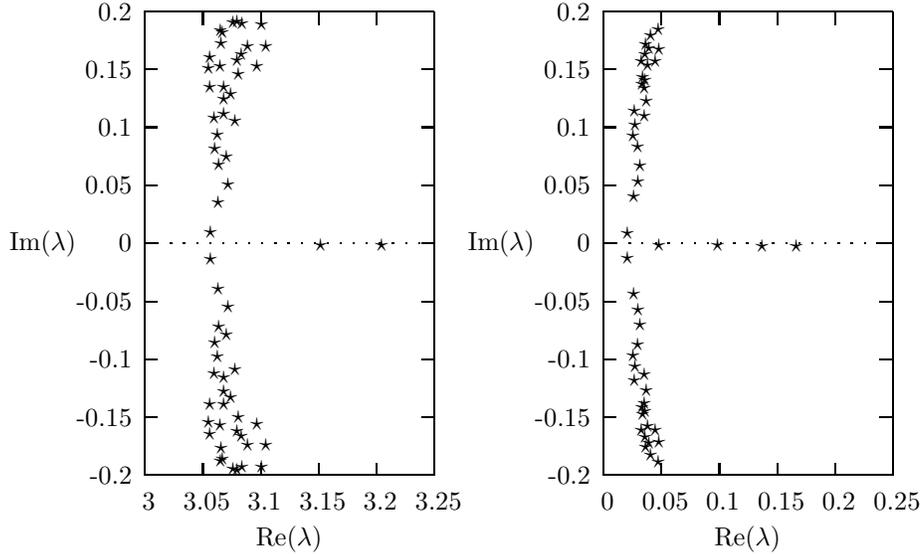

\centering{
\includegraphics[width=6cm]{figs/wi_spectrum_detail1.epsi}
\includegraphics[width=6cm]{figs/ci_conf100_runa.epsi}
}
\caption{Left: Part of the spectrum of \dwi\ 
for configuration nr. 100 of
run a. Right: The 50 smallest eigenvalues of \dci\ for the same
gauge-configuration.}
\label{spectrum_compare}
\end{figure}

Inspection of the spectrum of a Dirac operator is a very good method to
see how well chiral symmetry is approximated. In Fig. \ref{spectrum_compare},
for example, it can be clearly seen that the CI spectrum (right) deviates less
from the ideal Ginsparg-Wilson circle than that of \dwi\ (left).

\begin{figure}[ht]
\centering{\includegraphics[height=6.5cm]{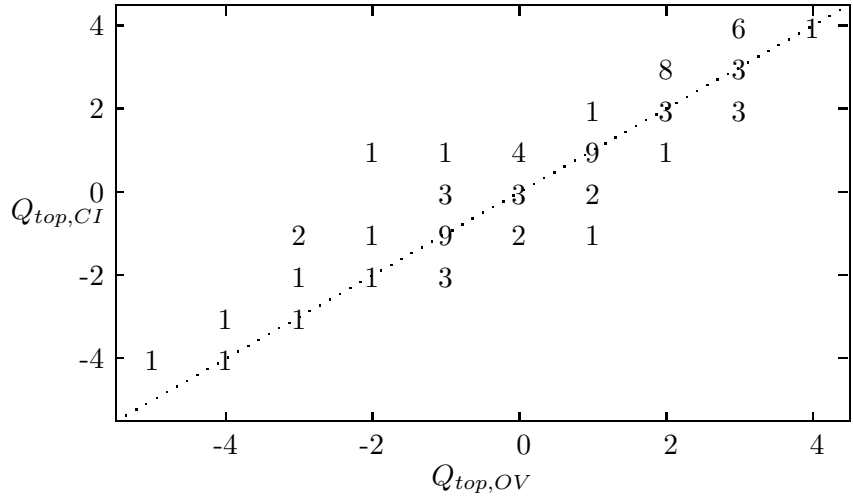}}
\caption{$Q_{top,OV}$ vs. $Q_{top,CI}$ for the gauge configurations
of run a. The numbers state how many configurations show a particular
combination of topological charges.}
\label{fig2}
\end{figure}

Another information the spectrum provides is the topological charge $Q_{top}$
of a gauge configuration. According to the Atiyah-Singer index theorem it is
possible to determine $Q_{top}$ by counting the zero modes of a
Ginsparg-Wilson type Dirac operator according to their chirality,
\begin{equation} 
Q_{top} = n_- - n_+ 
\end{equation}
with $n_\pm$ denoting the number of eigenmodes $|v\rangle$ with chirality
$\langle v\,|\,\gamma_5\,|\, v\rangle = \pm 1$ corresponding to eigenvalues
$\lambda=0$. When treating \dci, due to its approximate nature, these
eigenvalues are not exactly zero, but scatter on the real axis. For the CI
operator (and the Wilson operator) one may relate the number of real modes to
the topological sector. We  determine $Q_{top}$ by setting $n_\pm$ as the number
of eigenmodes $|w\rangle$ corresponding to real eigenvalues with  chirality
$\langle w\,|\,\gamma_5\,|\,w\rangle \gtrless 0$, respectively. (For eigenvalues
not on the real axis we numerically find $\langle w\,|\,\gamma_5\,|w\,\rangle =
0$ as expected.) When comparing the topological charge of a configuration
calculated with the exactly chiral \dov\ and the approximately chiral \dci, and
we find approximate agreement (cf. Fig. \ref{fig2}). The differences mostly
originate from missed eigenvalues far inside the Ginsparg-Wilson circle, which
are not recovered in our method of calculating eigenvalues, namely the program
package ARPACK; this tool only computes the eigenvalues with the smallest
absolute value with respect to a defined origin. When computing eigenmodes of
the overlap operator, the situation is different. There the sector depends on
the value of $s$ added to the diagonal part of the kernel operator (Wilson in
our case), which has frequent inner real modes that are missed in that overlap
projection, depending on $s$.

\begin{figure}[b]
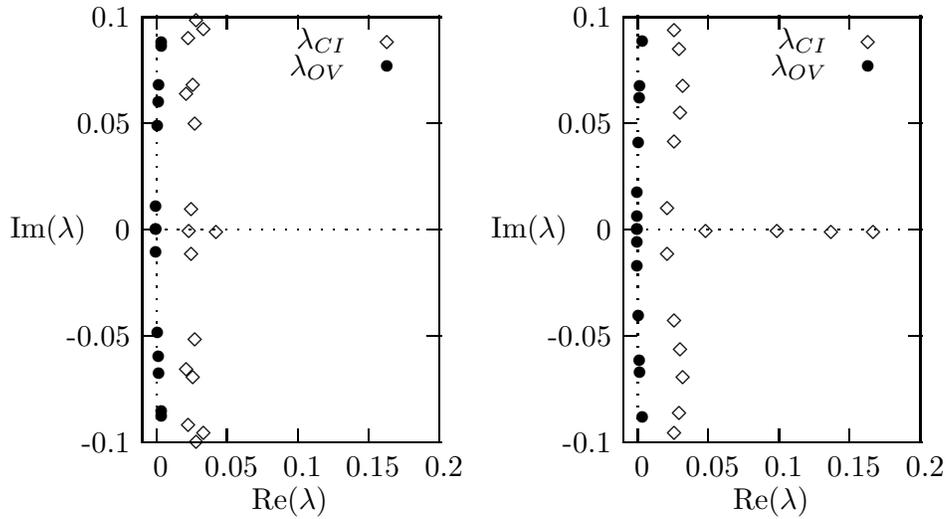

\begin{center}
\includegraphics[width=6cm]{figs/conf110.epsi}\hspace*{0.3cm}
\includegraphics[width=6cm]{figs/conf100.epsi}
\end{center}
\caption{Eigenvalues $\lambda_{OV}$ of \dov\ and eigenvalues 
$\lambda_{CI}$ of \dci\ in the complex plane. Left: configuration nr. 110 
(run (a)): $Q_{top}=-2$, $n_{+}=2$ for both
operators Right: configuration nr. 100 (run (a)):  $Q_{top}=-2$, but 
$n_{+,CI}=3$, $n_{-,CI}=1$.}
\label{ci_ov}
\end{figure}

If one does not take into account any normalization and directly compares
eigenvalues of \dci\ with those of \dov\ as defined in (\ref{dov}), an
interesting behavior can be seen. In cases where not only the topological charge
$Q_{top}$ is identical for both operators but also the number of real eigenvalues
$n_0$, the first few eigenvalues are in direct correspondence  (cf. Fig.
\ref{ci_ov}, left). However, if $n_0>Q_{top}$ for the CI operator, i.e., if
eigenmodes with opposite chirality cancel each other with respect to $Q_{top}$,
this correspondence is lost (cf. Fig. \ref{ci_ov}, right). When put into the
overlap operator, the real modes seem to ``move'' up resp. down the imaginary axis
(although not all of them close to the real axis),  thereby increasing the
eigenvalue density near the origin. For a similar observation  cf.
\cite{hasenfratz-2007}. This  enhancement leads to a higher value for the (bare)
chiral condensate when calculated directly with this definition of \dov.

\section{Localization of eigenmodes}

For the CI operator, we also calculated the inverse participation ratios ($ipr$)
of the eigenmodes for small eigenvalues, given by
\begin{equation}
ipr(\lambda) = V\;\sum_x \left(\sum_{\alpha,c} v(x,\alpha,c)^* v(x,\alpha,c)
\right)^2\;.
\end{equation}
The inner sum is over the color indices $c$ as well as the Dirac indices
$\alpha$, the outer sum over the space-time indices $x$. $V$ denotes the
space-time volume in lattice units,  and $v(x,\alpha,c)$ the eigenmode of \dci\
corresponding to the eigenvalue $\lambda$. This quantity is a good measure for
the localization properties of one eigenmode, with $ipr(\lambda)=1$ for a
non-localized mode and $ipr(\lambda)=V$ for a mode concentrated on only one
lattice point. To compute a suitable average over one HMC-run, we divided the
complex plane into a grid and calculated the average $ipr$ as 
\begin{equation}
\overline{ipr}=\frac{1}{n}\sum_{\lambda\in\Delta_\lambda} ipr(\lambda)\;,
\end{equation}
with $\Delta_\lambda$ being one square in the complex plane and $n$ the number of
eigenvalues in $\Delta_\lambda$. 

\begin{figure}
\begin{center}
\includegraphics[width=9cm]{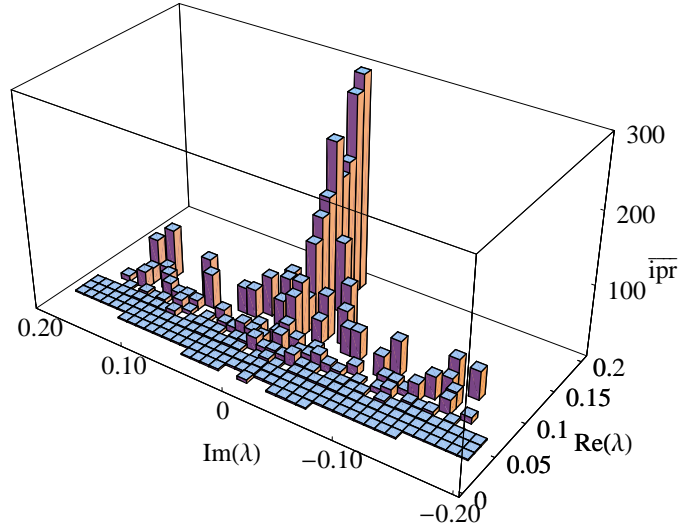}%
\end{center}
\caption{The average value $\overline{ipr}(\lambda)$ over the
complex plane for the  gauge configurations of run (a).}
\label{legoplot}
\end{figure}

In Fig. \ref{legoplot} we see, as expected, that $\overline{ipr}$ increases along
the real axis, and in general is higher for eigenvalues inside the
Ginsparg-Wilson circle (cf. also \cite{hasenfratz-2007}). For \dov\ one expects
$\overline{ipr}$ to be symmetric with respect to the real axis, but this is not
the case for \dci\ since it is not a normal matrix operator.

\section{Comparison with random matrix theory}

As the overlap operator implements chiral symmetry on the lattice exactly, the
distribution of the smallest eigenvalues (in leading order ChPT) is analytically
given by the well known results of random matrix theory (RMT) for the chiral
gaussian unitary ensemble \cite{Damgaard:2000ah}, at least in the
$\epsilon$-regime in the microscopic limit. Following the procedure in
\cite{Lang:2006ab}, where these calculations were done on the same configurations
for \dci, we thus compare our distributions for \dov\ to RMT. 

\begin{figure}[ht] \centering{
\includegraphics[width=5.5cm]{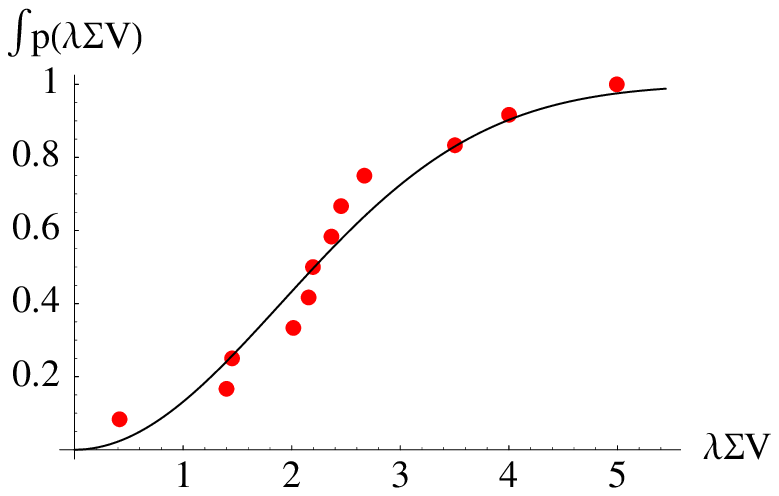}\hspace*{1cm}
\includegraphics[width=5.5cm]{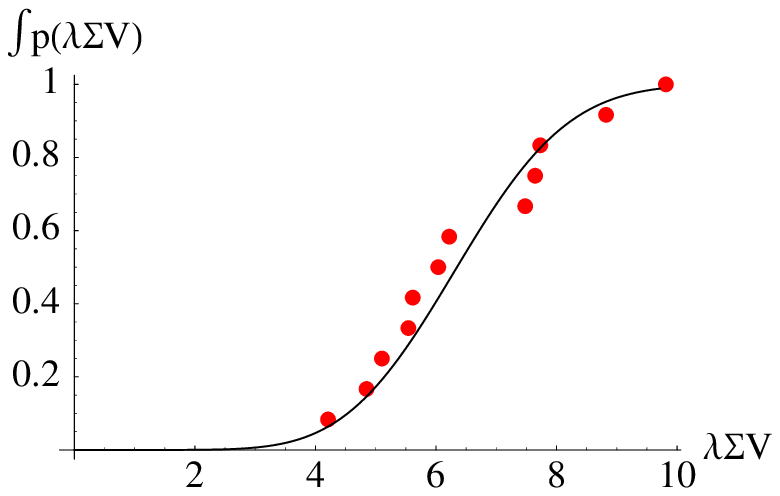}
\includegraphics[width=5.5cm]{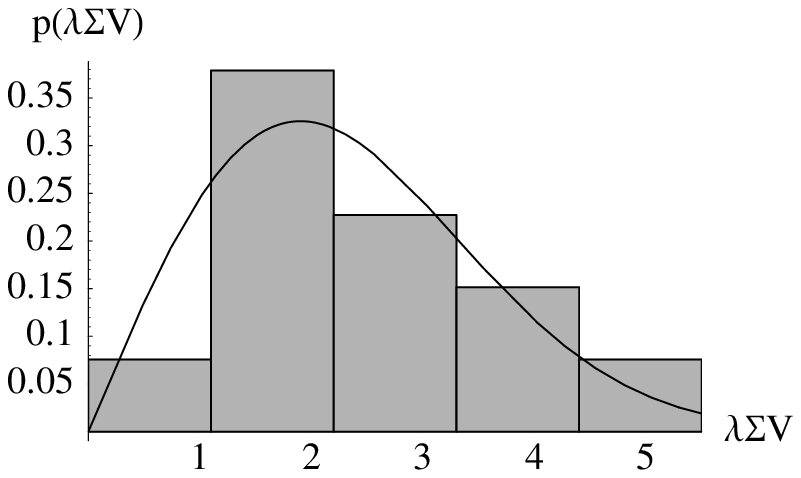}\hspace*{1cm}
\includegraphics[width=5.5cm]{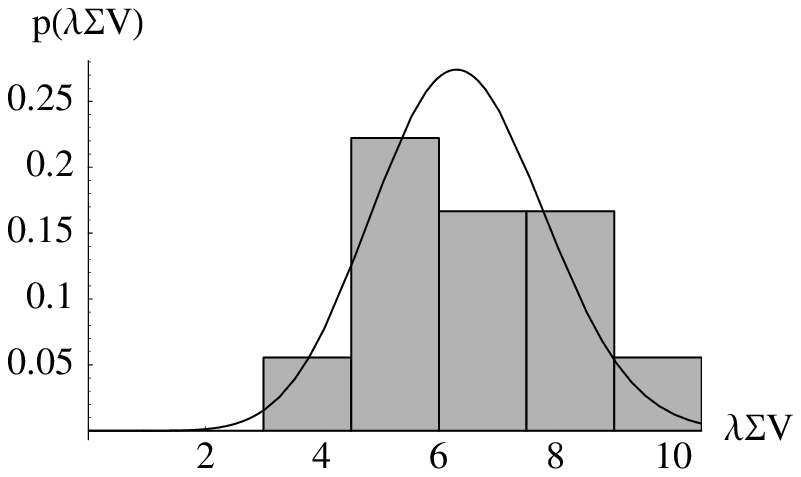} } \caption{The cumulative
distribution given by RMT compared to the distribution of the smallest
eigenvalues of the overlap operator. Left: $Q_{top}=0$, and k=1 (smallest
eigenvalue); Right: $Q_{top}=0$, and k=2 (second smallest eigenvalue), both for
run (a).} \end{figure}

A similar analysis for dynamical  overlap configurations was done in
\cite{DeLiSc06,fukaya-2007} and for 2-flavor staggered configurations in 
\cite{hasenfratz-2006-LAT2006}. For the fits that determine the chiral condensate
$\Sigma$, we use the cumulative eigenvalue distributions and look at the smallest
and second smallest eigenvalue in the topological sectors $Q_{top}=0$ and
$|Q_{top}|=1$. 

The values for $\Sigma$ for run (a)--(d), determined from the spectra of \dov,
are given in Table \ref{sigma}. The fit was done using the Kolmogorov-Smirnov
test, the errors computed via statistical bootstrap. 

\begin{table}[t]
\centering{\small
\begin{tabular}{|c|c|c|c|c|}
\hline
HMC run & $k$ & $|Q_{top}|$&\#confs. & $-(\Sigma)^{1/3}$ $\rm{MeV^{1/3}}$\\
\hline
a& 1 & 0&12 &338(6)\\
& 1 & 1&25&332(3)\\
&2&0&12&310(3)\\
&2&1&25&319(3)\\
\hline
b&1&0&7&353(10)\\
&1&1&8&350(9)\\
&2&0&7&362(5)\\
&2&1&8&330(6)\\
\hline
c&1&0&17&350(6)\\
&1&1&12&346(9)\\
&2&0&17&340(3)\\
&2&1&12&322(5)\\
\hline
d&1&0&5&365(18)\\
&1&1&11&370(10)\\
&2&0&5&348(14)\\
&2&1&11&346(2)\\
\hline
\end{tabular}}
\caption{Results for the value of the bare condensate $\Sigma$,
for all runs of the HMC. The value of $k$ refers to the smallest (1) or
second smallest (2) (imaginary part of the) eigenvalue.}
\label{sigma}
\end{table}

\section{Problems and issues}

The resulting values for the condensate still have to be renormalized. This can
be done by determining the renormalization constant $Z_S$ by standard tools of
non-perturbative renormalization. In the case of \dci\ the renormalization
constants for the dynamical case have been computed in recent work \cite{huber}.
For \dov, the situation is more involved, and the renormalization depends on
the value of $s$. The weak coupling expansion for small momenta $p$ has a
behavior $\dov = i\;\gamma_\mu p_\mu/s+\mathcal{O}(p^2)$. This changes for the
interacting case in a  non-trivial way. Comparing the results for the bare
condensate for CI \cite{Lang:2005jz} with overlap, we expect $Z_{S,OV}=Z_{S,CI}\,
\Sigma_{CI}/\Sigma_{OV}\approx 0.64 Z_{S,CI}$.

The dependence on the physical quark mass is also worth being explored. To this
point we assume that the AWI-mass computed for our dynamical gauge fields is the
same as the quark mass entering the RMT-fits. In \cite{hasenfratz-2006-LAT2006}
the overlap quark mass has been determined using the distribution of topological
charge, and found to be different from the value calculated for the underlying
gauge fields. We do not have sufficiently high statistics to follow this approach,
but nevertheless the sea quark mass of the overlap operator should be adjusted,
such that the pion mass (or the AWI-mass) computed with the overlap operator
agree with the pion mass of the CI operator.

\acknowledgments
Fruitful discussions with C. Gattringer, A. Hasenfratz and P. Majumdar are
gratefully acknowledged. M.J. wants to thank the Paul Urban-Stiftung for
financial support.

\end{document}